\documentclass[a4paper, 10pt, conference]{ieeeconf}      

\IEEEoverridecommandlockouts                              

\usepackage{amsmath} 
\usepackage{amssymb}  
\usepackage{cite}
\usepackage{graphicx}
\usepackage[caption=false, font=footnotesize]{subfig}

\usepackage{tikz}
\usepackage{pgfplots}
\usepgfplotslibrary{fillbetween}

\newcommand\copyrighttext{%
	\footnotesize Copyright $\copyright$ 2019 IEEE.
	Personal use of this material is permitted.
	Permission from IEEE must be obtained for all other uses, in any current or future media, including reprinting/republishing this material for advertising or promotional purposes, creating new collective works, for resale or redistribution to servers or lists, or reuse of any copyrighted component of this work in other works.}%
\newcommand\copyrightnotice{%
	\begin{tikzpicture}[remember picture,overlay]%
	\node[anchor=south,yshift=10pt] at (current page.south) {\fbox{\parbox{\dimexpr\textwidth-2cm}{\copyrighttext}}};%
	\end{tikzpicture}%
	\vspace{-10pt}%
}

\title{\LARGE \bf
LACI: Low-effort Automatic Calibration of Infrastructure Sensors$^\ast$
}

\author{Johannes M{\"u}ller$^{1}$, Martin Herrmann$^{1}$, Jan Strohbeck$^{1}$, Vasileios Belagiannis$^{1}$ and Michael Buchholz$^{1}$
	\thanks{*Part of this work was financially supported by the Federal Ministry of Economic Affairs and Energy of Germany within the program "Highly and Fully Automated Driving in Demanding Driving Situations" (project MEC-View, grant number 19A16010I). Another part of this work has been conducted as part of ICT4CART project which has received funding from the European Union’s Horizon 2020 research \& innovation program under grant agreement No. 768953. Content reflects only the authors’ view and European Commission is not responsible for any use that may be made of the information it contains. }
	\thanks{$^{1}$Johannes M{\"u}ller, Martin Herrmann, Jan Strohbeck, Vasileios Belagiannis and Michael Buchholz are with the Institute of Measurement, Control and Microtechnology,
		Ulm University, D-89081 Ulm, Germany
		{\tt\small \{johannes-christian.mueller, martin.herrmann, jan.strohbeck, vasileios.belagiannis, michael.buchholz\}@uni-ulm.de}
	}%
}

\begin{document}

\maketitle
\copyrightnotice
\thispagestyle{empty}
\pagestyle{empty}

\begin{abstract}

Sensor calibration usually is a time consuming yet important task. While classical approaches are sensor-specific and often need calibration targets as well as a widely overlapping field of view (FOV), within this work, a cooperative intelligent vehicle is used as callibration target. The vehicle is detected in the sensor frame and then matched with the information received from the cooperative awareness messages send by the coperative intelligent vehicle. The presented algorithm is fully automated as well as sensor-independent, relying only on a very common set of assumptions. Due to the direct registration on the world frame, no overlapping FOV is necessary. The algorithm is evaluated through experiment for four laserscanners as well as one pair of stereo cameras showing a repetition error within the measurement uncertainty of the sensors. A plausibility check rules out systematic errors that might not have been covered by evaluating the repetition error.

\end{abstract}

\section{INTRODUCTION}
The sensor calibration of a traffic monitoring system usually is a time consuming yet important task \cite{Datondji_Survey2016}. While intrinsic calibration of the sensor setup is often either already done by the sensor manufacturer or can be performed in the lab, the extrinsic calibration is specific to the respective traffic monitoring system and has to be performed on the field.

Classical approaches are sensor-specific, e.g. \cite{Meissner2010} for multiple laser scanners and \cite{Datondji2017} for a wide baseline fisheye-stereo setup, and often require specific calibration objects, e.g. \cite{Meissner2010} using a specific polygon or \cite{Kuemmerle2018} using a sphere, that need to be moved through the scene. Thus, the road under observation has to be closed for several hours during calibration. This means a lot of effort, as the calibration has to be repeated whenever a sensor slightly moves, e.g. during a thunderstorm. Furthermore, most calibration schemes register the sensors only relatively to each other and therefore need an widely overlapping field of view (FOV), e.g. \cite{Datondji2017}, \cite{Kuemmerle2018}. This, however, does not fit well with the requirements of traffic monitoring, where a maximum FOV should be covered with a minimum number of sensors \cite{Yin2015}.

We address the problem by detecting the bounding box of a cooperative intelligent vehicle (CIV) in the sensor frame and then matching it to the corresponding bounding box in the world frame as communicated by the CIV, e.g. via Cooperative Awareness Messages (CAM) \cite{ETSI_CAM2011}. Corresponding bounding boxes are associated with respect to their timestamps and the calibration result is improved by running a regression over multiple matched bounding boxes. The matching of the bounding boxes recovers the affine transform from the sensor frame into the world frame which in fact is the extrinsic calibration of the respective sensor.

As modern automated vehicles can localize themselves with high accuracy \cite{Stuebler2015}, we are able to reach repetition errors within the sensor accuracy of the traffic monitoring system (see Section \ref{sec:Evaluation}). By matching bounding boxes, which can be detected in all common sensors used for traffic monitoring  \cite{Datondji2017} that feature a 3D point cloud, the proposed method functions sensor-independently. Furthermore, because the CIV is already localized in the world frame, each sensor can be registered to the world frame individually, thus no overlapping FOV is needed.

The key contributions of this paper can be summarized as follows:
\begin{itemize}
	\item We propose a self-calibration scheme that is significantly easier and faster compared to placing calibration objects in the scene.
	\item The self-calibration algorithm is fully automated.
	\item The algorithm is robust towards static background objects and can identify them for future use.
	\item By applying the algorithm to a multi-sensor setup consisting of four laserscanners and one stereo camera, the sensor-independence of the algorithm is demonstrated.
	\item The algorithm is evaluated through experiment showing a repetition accuracy within the measurement accuracy of the sensors.
\end{itemize}

\subsection{Related Work}
Calibration has been an ongoing topic in the recent years. In 2016, Datondji et al. \cite{Datondji_Survey2016} have determined a major challenge in traffic monitoring to develop automatic calibration methods and a general pipeline for traffic monitoring at intersections and see a possible future major contribution to the challenge to directly exploit vehicles motion for calibration. In a following paper of that research group \cite{Datondji2017}, a vehicle driving through the scene has been used as calibration target to calibrate a pair of fish-eye stereo cameras to each other. While promising results are shown, their method is bounded to the sensor type. Furthermore, for their stereo vision application, naturally, their cameras need to have a widely overlapping FOV, which is not the case for general traffic monitoring systems.

A plane-based calibration scheme for a multi-camera setup with non-overlapping FOV has been presented in \cite{Zhu2016}, showing high calibration accuracy. While, as promised, the method does not need an overlapping FOV for \textit{all} cameras, a pairwise widely overlapping FOV as well as textured environment is still needed. In contrast, the method presented within this work does not even require a pairwise overlapping FOV.

Finally, in \cite{AtaerCansizoglu2014}, multiple cameras with non-overlapping field of views are calibrated using a simultaneously localization and mapping (SLAM) technique of a mobile robot. While in \cite{AtaerCansizoglu2014} actually the map resulting from the SLAM is used to do the extrinsic calibration, in our work, only the localization result is used. Thus, only the world coordinates must be transmitted rather than the whole map, which would be infeasible for cooperative intelligent transportation system applications due to bandwidth limitations.

The rest of the paper is structured as follows: In Section \ref{sec:Algorithm}, the calibration algorithm is described in detail, followed by the experimental evaluation shown in Section \ref{sec:Evaluation}. Finally, conclusions are drawn in Section \ref{sec:conclusion}.

\section{SELF-CALIBRATION ALGORITHM} \label{sec:Algorithm}
In this section, the calibration algorithm is described. First of all, the problem formulation, as well as the necessary assumptions are stated. Then, starting from an algorithm overview, the calibration algorithm is presented in detail.

\subsection{Problem Formulation}
Consider a multi-type-multi-sensor network consisting of $N_S$ sensors $S_i, \, i \in [1,N_S]$, for which the intrinsic parameters are assumed to be known\footnote{We further assume that the stereo parameters resulting from the registration of a pair of mono-cameras are known. These can be obtained using a state-of-the-art auto-calibration mechanism for stereo, e.g. \cite{Moghadam2013}. }.
Then, each point $\boldsymbol{p}_{S_i}$ in the homogeneous sensor coordinate frame can be mapped into the homogeneous world coordinate frame via the linear transform $\mathcal{T}^{w}_{S_i}$ by
\begin{equation}
	\label{eq:Problem}
	\boldsymbol{p}_w = \underbrace{\begin{bmatrix}
	\boldsymbol{R}^{w}_{S_i} & \boldsymbol{t}^{w}_{S_i} \\
	\boldsymbol{0} & 1
	\end{bmatrix}}_{\mathcal{T}^{w}_{S_i}} \; \cdot \; \boldsymbol{p}_{S_i} \; ,
\end{equation}
where $\boldsymbol{R}^{w}_{S_i}$ is a rotation matrix reflecting the three rotatory degrees of freedom (DOFs), while $ \boldsymbol{t}^{w}_{S_i}$ is a translation vector representing the translatory DOFs. A world point is given as $\boldsymbol{p}_w = [ x, \; y, \; z, \; 1]^{T}_w$, where $x$ and $y$ are the north and east Universal Transverse Mercartor (UTM) coordinates, respectively, and $z$ is the height over ground.
The goal of the self-calibration algorithm is to recover the linear transform $\mathcal{T}^{w}_{S_i}$ of (\ref{eq:Problem}), i.e. the extrinsic calibration.

In order to reach that goal, the algorithm requires the following assumptions:
Each sensor provides a 3D point cloud that is sufficiently dense to detect the CIV within it. Furthermore, the area under supervision has to be plane in first proximity.
To ensure that the Karhunen-Loeve transform \cite{Friedman2001} used to fit a plane through the ground points, finds good enough principle components, there must be significantly more ground points than points originating from static background elements. Our own, as well as others' \cite{Meissner2010} experience shows that this assumption holds in practice.
To extract the static background elements, a background measurement is needed that shows the scene without any road user. In a second measurement, the CIV needs to be the only vehicle in the respective region of interest (ROI).
Note that these measurements are done within seconds, so a road closure can be avoided by using appropriate time slots. For online re-calibration, the requirements are significantly relaxed, as the ROI for ground plane detection as well as the expected position of the CIV are roughly known. Thus, the respective ROI is significantly smaller than the sensor FOV. Finally, the CIV and the sensors need a common time base, e.g. from a global navigation satellite system.

\subsection{Algorithm Overview}
\begin{figure}[bth]
	\centering
	\includegraphics[width=0.4\textwidth]{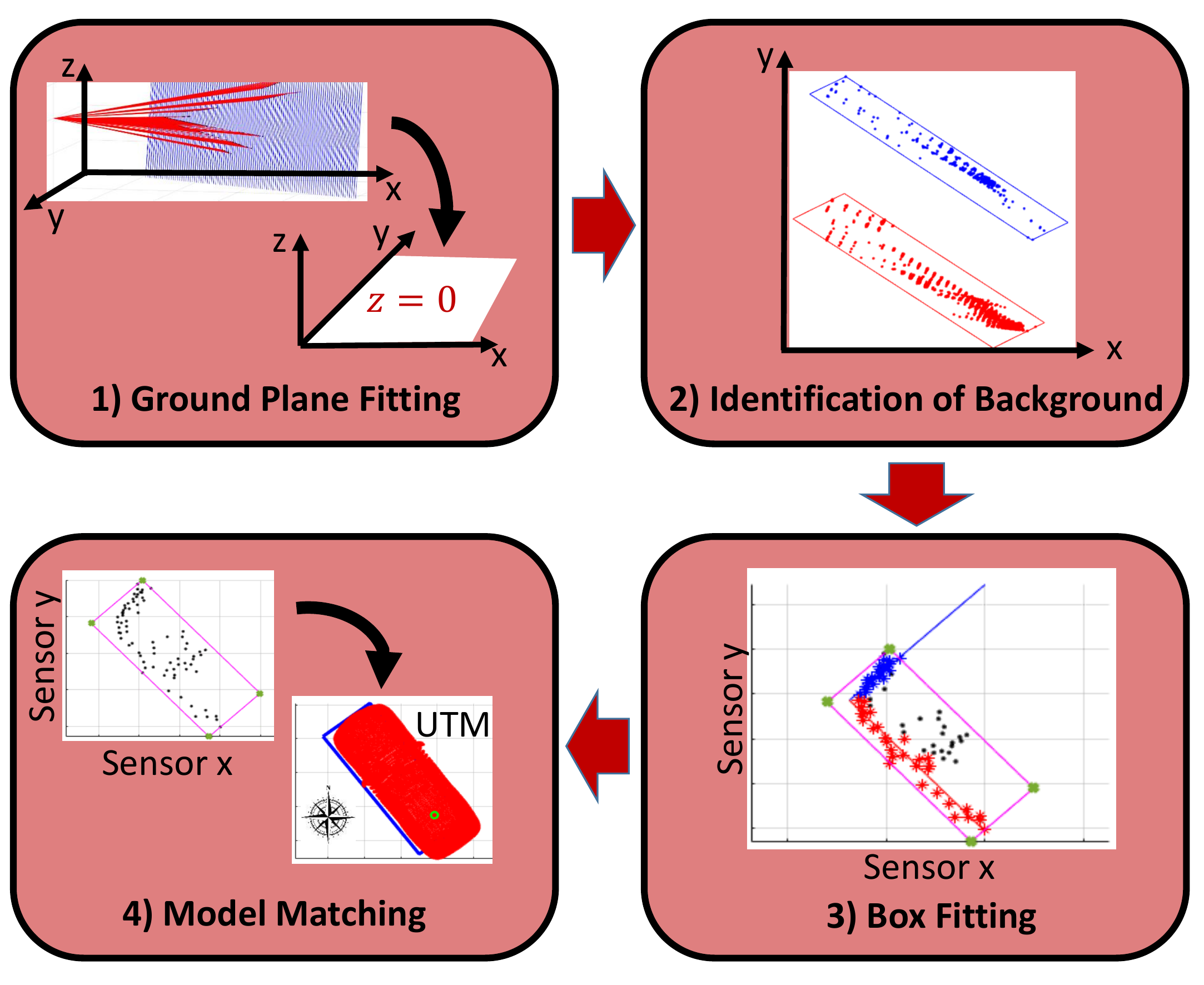}
	\caption{Algorithm Overview. The four basic steps for finding the extrinsic calibration are illustrated.}
	\label{fig:Overview}
\end{figure}

 The key idea of the algorithm is to first fit a ground plane into the sensor frame to reduce the 3D problem to 2D by omitting the height. Then, the remaining DOFs are resolved by matching a bounding box detected within the sensor frame to the corresponding bounding box of the CIV in the world frame. The algorithm therefore performs the following steps:

\begin{enumerate}
	\item Ground plane fitting.
	\item Automatic identification of background elements.
	\item Box fitting of the target vehicle.
	\item Model matching between the detected vehicle in the sensor frame and the vehicle's position in world coordinates.
\end{enumerate}
The algorithm steps are sketched in Fig. \ref{fig:Overview}. The following subsections explain the respective steps in detail.

\subsection{Ground Plane Fitting} \label{sec:GroundPlaneFit}
The basic idea of the ground plane fitting is to extract the ground points from the background measurement (i.e. the measurement of an empty scene), fit a plane through it and then rotate and move the fitted plane such that the fitted plane matches the $z=0$ plane of the world coordinate system, which indicates the surface of the road as recorded in the map. Thus, for sloping roads, the slope has to be compensated for during the projections later on.

First, the plane is fitted by applying the Karhunen-Loeve transform \cite{Friedman2001} to the points $\boldsymbol{p}_i$ of the incoming point cloud $\boldsymbol{P} \in \mathbb{R}^{3 \times N}$,

\begin{equation}
		\small
		\boldsymbol{V} \boldsymbol{\Sigma} \boldsymbol{N}^{T} = \text{svd}\left\{ \boldsymbol{P} -  \frac{1}{N}\sum_{i=1}^{N} \boldsymbol{p}_i \right\} \; , \\
		\boldsymbol{V} =  [\boldsymbol{v}_1 \; \boldsymbol{v}_2 \; \boldsymbol{n}] \; ,
\end{equation}
where $N$ is the number of points in the input point cloud, $\text{svd}\{ \cdot \} $ is the singular value decomposition, $\boldsymbol{V}$ is the matrix containing the principle components $\boldsymbol{v}_1$, $\boldsymbol{v}_2$, and $\boldsymbol{n}$, and $\boldsymbol{\Sigma}$ and $\boldsymbol{N}$ are the matrices containing the singular values and the right-singular vectors. Due to the assumption that the ground points significantly outnumber the points originating from the static background elements (which can be enhance by a rogh ROI prior), $\boldsymbol{v}_1$ and $\boldsymbol{v}_2$ are base vectors of the ground plane, while $\boldsymbol{n}$ is the normal vector to the ground plane.
The mounting height of the sensor then can be determined by
\begin{equation}
	\small
	\hat{h} = - \boldsymbol{V}^{-1} \cdot (\frac{1}{N}\sum_{i=1}^{N} \boldsymbol{p}_i) \; ,
\end{equation}
so that the overall mapping
\begin{equation}
	\small
	\mathcal{T}_{S_i}^{\tilde{S}_i} = \begin{bmatrix}
		\boldsymbol{V}^{-1} & \begin{matrix}
		0 \\ 0 \\ \hat{h}
		\end{matrix} \\
		0 \; 0 \; 0 & 1
	\end{bmatrix}
\end{equation}
transforms the fitted ground plane to the $z=0$ plane.

A priori, the algorithm has no knowledge which of the points are actually ground points and which belong to static background elements. Hence, in the first run, the algorithm fits the ground plane through all available points. However, due to the assumption that the ground points in an empty scene significantly outnumber the points originating from a background element, it can be concluded that the first estimation of the ground plane will be close to the real ground plane. Thus, all points that are far away from the estimated ground plane can be identified as originating from background elements other than the ground plane and therefore can be ignored while running a second ground plane fit. This results in a far better estimation of the ground plane and makes the algorithm robust towards static background elements in the scene. This procedure can be repeated iteratively, till the result is sufficiently good in terms of e.g. the least squares error.

\subsection{Automatic Identification of Background Elements} \label{sec:BackgroundSubtraction}
First of all, the ground points are filtered out from a background measurement
by applying a threshold $\theta_g$ to the $z$-component of the measurement points after the ground plane fitting step $ \boldsymbol{P}_{\tilde{S}_i}$. This yields
\begin{subequations}
	\begin{align}
		\mathcal{S}_{\text{ground}} &= \{ \boldsymbol{p}_{\tilde{S}_i} \in \boldsymbol{P}_{\tilde{S}_i} | \boldsymbol{p}_{z, \, S_i} \leq \theta_g \} \\
		\mathcal{S}_{\text{stat}} &= \{ \boldsymbol{p}_{\tilde{S}_i} \in \boldsymbol{P}_{\tilde{S}_i} | \boldsymbol{p}_{z, \, \tilde{S}_i} > \theta_g \} \; ,
	\end{align}
\end{subequations}
where $\mathcal{S}_{\text{ground}}$ is the set of all ground points, while $\mathcal{S}_{\text{stat}}$ is the complement set consisting of all points above the ground which are considered to belong to static background objects. To make the segmentation of the ground points robust against measurement noise and unevenness of the ground, here $\theta_g = 0.5 \, \text{m}$ is chosen slightly above the ground plane.

However, the representation of background elements as set of points is not very handy yet, as for each incoming point, the distance to all points in $\mathcal{S}_{\text{object}}$ would have to be calculated to classify whether the incoming points belong to the foreground or background.

Thus, $\mathcal{S}_{\text{stat}}$ is fed into the box fitting (see Section \ref{sec:BoxFitting}) to estimate the bounding boxes around the background elements. These bounding boxes transformed to the world frame then can be used for online object detection. 

\subsection{Box Fitting} \label{sec:BoxFitting}
In the next step, another measurement, this time including the CIV, is taken. The CIV's shape is extracted from the point cloud, and a bounding box is fitted around it. Therefore, first of all, the input point cloud is transformed to the $\tilde{\mathcal{S}}_i$ frame.
The ground points points are removed by  applying the threshold $\theta_g$, and a point-in-polygon test \cite{Hormann2001} in combination with the detected background boxes (see Section \ref{sec:BackgroundSubtraction}) is applied to remove the static background elements. In order to make the algorithm more robust against clutter as well as residual background elements, the remaining points are clustered with the DBSCAN algorithm \cite{Ester1996DBSCAN} and all but the points from the cluster with the highest number of points are discarded. The remaining set of points $\mathcal{S}_{\text{vehicle}}$ is used for further processing.

Through the ground plane fitting, three of the six DOFs are already determined. Hence, the problem can be reduced to 2D by projecting all points to the $z=0$ plane, which significantly reduces complexity for the following steps.

Depending on the sensor type and the direction under which the CIV is observed, the vehicle appears as L-, I- or box shape within the projected 2D points $\mathcal{S}_{\text{vehicle,2D}}$. Box shape fitting algorithms are usually based on the Karhunen-Loeve transform, thus, if the assumed box shape actually is an L-shape, these class of algorithms fail. In turn, a box shape can easily be reduced to an I-shape and an I-shape can be considered as special case of an L-shape, where loosely speaking the bar of the L is missing. Hence, an L-shape fitting algorithm is used to perform the box fitting.

For the L-shape fitting, we settle upon the idea of Shen et al. \cite{Shen2015}, who first partition the input points into two sets $\mathcal{P}$ and $\mathcal{Q}$, $ \mathcal{P} \bigcup \mathcal{Q} = \mathcal{S}_{\text{vehicle,2D}}$ and then fit two lines
\begin{subequations}
	\begin{align}
		\mathcal{L}_{\mathcal{P}} &: \{ (x_P,y_P)_{i=1 \dots p} \in \mathcal{P}: \alpha_1 x_P + \alpha_2 y_P + c_1 = 0 \}  , \\
		\mathcal{L}_{\mathcal{Q}} &: \{ (x_Q,y_Q)_{i=1 \dots q}  \in \mathcal{Q}: \alpha_3 x_Q + \alpha_4 y_Q + c_2 = 0\}
	\end{align}
\end{subequations}
with respect to the least squares error, which fulfill the constraint that $\mathcal{L}_P$ and $\mathcal{L}_Q$ must be orthogonal. This is achieved by solving the optimization problem
\begin{subequations}
	\label{eq:OptimizationProblem}
	\begin{align}
	\small
		\boldsymbol{u} &= \arg \min_{\boldsymbol{u}} \left\{ \lVert \boldsymbol{A} \boldsymbol{u} \rVert \right\} \text{s. t.} \label{eq:OptimizationProblemJ}\\
		\boldsymbol{A}^T &= \begin{bmatrix}
						1 & \cdots & 1 & 0 & \cdots & 0\\
						0 & \cdots & 0 & 1 & \cdots & 1\\
						x_{P,1} & \cdots & x_{P,p} & y_{Q,1} & \cdots & y_{Q,p}\\
						y_{P,1} & \cdots & y_{P,p} & -x_{Q,1} & \cdots & -x_{Q,p}
					   \end{bmatrix}\;,\\ 
		\boldsymbol{u} &= \begin{bmatrix}
						c_1 & c_2 & \alpha_1 & \alpha_2
					  \end{bmatrix}^{T} = \begin{bmatrix} \boldsymbol{c} & \boldsymbol{\alpha} \end{bmatrix} \, , \\
				 1 &= \left \lVert \begin{bmatrix}
						0 & 0 & 1 & 0 \\
						0 & 0 & 0 & 1
					  \end{bmatrix}	\, \cdot \, \boldsymbol{u} \right \rVert	\, .	  					  			 
	\end{align}
\end{subequations}
The optimization problem (\ref{eq:OptimizationProblem}) can be transformed into the optimization problem
\begin{equation}
	\label{eq:TransforemdOptimizationProblem}
	\small
	\boldsymbol{\alpha} = \arg \min_{\boldsymbol{\alpha}} \Big\{ \boldsymbol{\alpha}^{T} \underbrace{(\boldsymbol{M}_{22} - \boldsymbol{M}_{12}^{T} \boldsymbol{M}_{11}^{-1} \boldsymbol{M}_{12})}_{=: \tilde{\boldsymbol{M}}} \boldsymbol{\alpha} \Big\} \text{ s.\,t. } \lVert \boldsymbol{\alpha} \rVert = 1 \, ,
\end{equation}
where
\begin{equation}
	\small
	\boldsymbol{M} = \boldsymbol{A}^{T} \boldsymbol{A} = \begin{bmatrix}
	\boldsymbol{M}_{11} & \boldsymbol{M}_{12} \\
	\boldsymbol{M}_{21} & \boldsymbol{M}_{22}
	\end{bmatrix} \, \text{, } \boldsymbol{M}_{ii} = \in \mathbb{R}^{2 \times 2}\, ,
\end{equation}
and $||\cdot||$ is the $l_2$ vector norm.
The transformed optimization problem (\ref{eq:TransforemdOptimizationProblem}) then can be calculated by computing the eigenvectors of $\tilde{\boldsymbol{M}} \in \mathbb{R}^{2 \times 2}$, as the eigenvector corresponding to the smallest eigenvalue (i.e. singular vector of $\boldsymbol{A}$) minimizes the cost function of (\ref{eq:OptimizationProblemJ}). Finally, $\boldsymbol{c}$ can be estimated by
\begin{equation}
	\boldsymbol{c} = \boldsymbol{M}_{11}^{-1} \boldsymbol{M}_{12} \, \boldsymbol{\alpha} \, .
\end{equation}

While Shen et al. have no method for partitioning $\mathcal{S}_{\text{vehicle,2D}}$ other than brute forcing the partitioning using some heuristics specific to laserscanners, we improve upon the L-shape fitting algorithm of Shen et al. by adding a partitioning strategy.

The intuition behind the strategy is sending wide beams of light onto a field of particles (the points in the point set) and estimating the orientation of the particle field by running a linear regression on the reflections closest to the light source.

For efficient implementation, the imaginary light source in general is oriented in parallel with one coordinate axis, while, without loss of generality, we assume here that the light is sent in positive y direction. The partitioning then consists of the following steps: First of all, $x_\text{min} = \min_x\{ \mathcal{S}_{\text{vehicle,2D}} \}$ and $x_\text{max} = \max_x\{ \mathcal{S}_{\text{vehicle,2D}} \}$ are determined and the cluster points are grouped with respect to their x-coordinate
\begin{equation}
	\small
	\mathcal{S}_{P_k} =  \{ \boldsymbol{p}_j \in \mathcal{S}_{\text{vehicle,2D}} : x_{p_j} \in [x_{\text{min}} + (k-1) \Delta x, \, x_{\text{min}} + k \Delta x ]  \} ,
\end{equation}
where $k \in \mathbb{N}$ and $\Delta x$, figuratively the width of the light beams, is a parameter reflecting the density of the point cloud. Then, a linear regression on the set of points
\begin{equation}
	\mathcal{S}_{y,\text{min}} = \{ \boldsymbol{p}_{k,y \text{min}} \in \mathcal{S}_{P_k} : y_{p_{k,y \text{min}}} = \min_{y} \{ \mathcal{S}_{P_k} \} \}
\end{equation}
corresponding to the closest reflections is calculated, resulting in the line $\mathcal{L}_r: \{ y_r = a x_r + b \}$. In the next step, all cluster points are projected to $\mathcal{L}_r$ through
\begin{equation}
	\small
	\tilde{\mathcal{S}}_{\text{vehicle,2D}} = \left\{ \begin{bmatrix}
	1 & -a \\
	a & -1
	\end{bmatrix}^{-1} \begin{bmatrix}
	x \\
	y - b
	\end{bmatrix} : \; \begin{bmatrix}
	x \\
	y
	\end{bmatrix} \in \mathcal{S}_{\text{vehicle,2D}}  \right\} \, ,
\end{equation}
and a histogram is calculated on $\tilde{\mathcal{S}}_{\text{vehicle,2D}}$ with respect to its $x$-coordinate. For an ideal L-shape and a perfect orientation estimation, all points forming the bar of the L would be projected onto a single point on $\mathcal{L}_r$. Under non-ideal conditions, the points corresponding to the bar are still expected to form a very dense region on $\mathcal{L}_r$. Hence, the subset of points in $\tilde{\mathcal{S}}_{\text{vehicle,2D}}$ corresponding to the histogram bin with the highest count are assigned to $\mathcal{P}$, while all point within a tolerance band of $\theta_\text{thresh}$ around $\mathcal{L}_r$ are assigned to $\mathcal{Q} = \{ \tilde{\boldsymbol{p}} \in \tilde{\mathcal{S}}_{\text{vehicle,2D}} : | y_{\tilde{p}}| \leq \theta_\text{thresh} \}$.

\begin{figure}[bt]
	\centering
	\begin{tikzpicture}
	\footnotesize
	\input{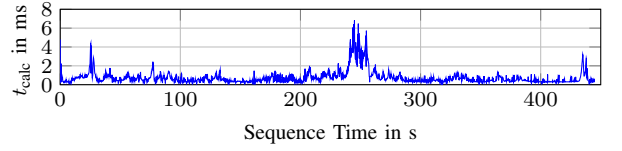}
	\end{tikzpicture}	
	\caption{Evaluation of the box detection for the four laserscanners. The sampling time is $80 \, \text{ms}$, hence, with an average calculation time of $\text{mean}(t_\text{calc}) \approx 2 \, \text{ms}$, the algorithm is real-time capable.}
	\label{fig:Runtime}
\end{figure}

The partitioning strategy presented within this paper is particularly beneficial if a large number of points, e.g. from a stereo vision system, has to be processed in real time. Figure \ref{fig:Runtime} shows an evaluation of the run time consumed by the processing of the four laserscanners (cf. Section \ref{sec:Setup}) on a Intel Xeon E5-1630v4 processor. The typical calculation time is at about  $t_\text{calc} \approx 2 \, \text{ms}$ which is $2.5\,\%$ of the sampling time $T_S = 80 \, \text{ms}$.

However, in contrast to object detection tasks, where the presented box fitting step can be also used for, the calculation time is of minor priority for the calibration. Nevertheless, the partitioning strategy presented within this work is still advantageous compared to the grid search proposed by Shen et al, as it is more flexible towards the vehicle's appearance.
  If an I- or box shape is detected instead of an L-shape, this will be reflected within the histogram, as multiple histogram bins would show approximately the same count. In this case, $p=1$ can be set such that the orientation is purely estimated based on the regression over $\mathcal{Q}$.

\begin{figure}[bth]
	\centering
	\includegraphics[width=0.37\textwidth]{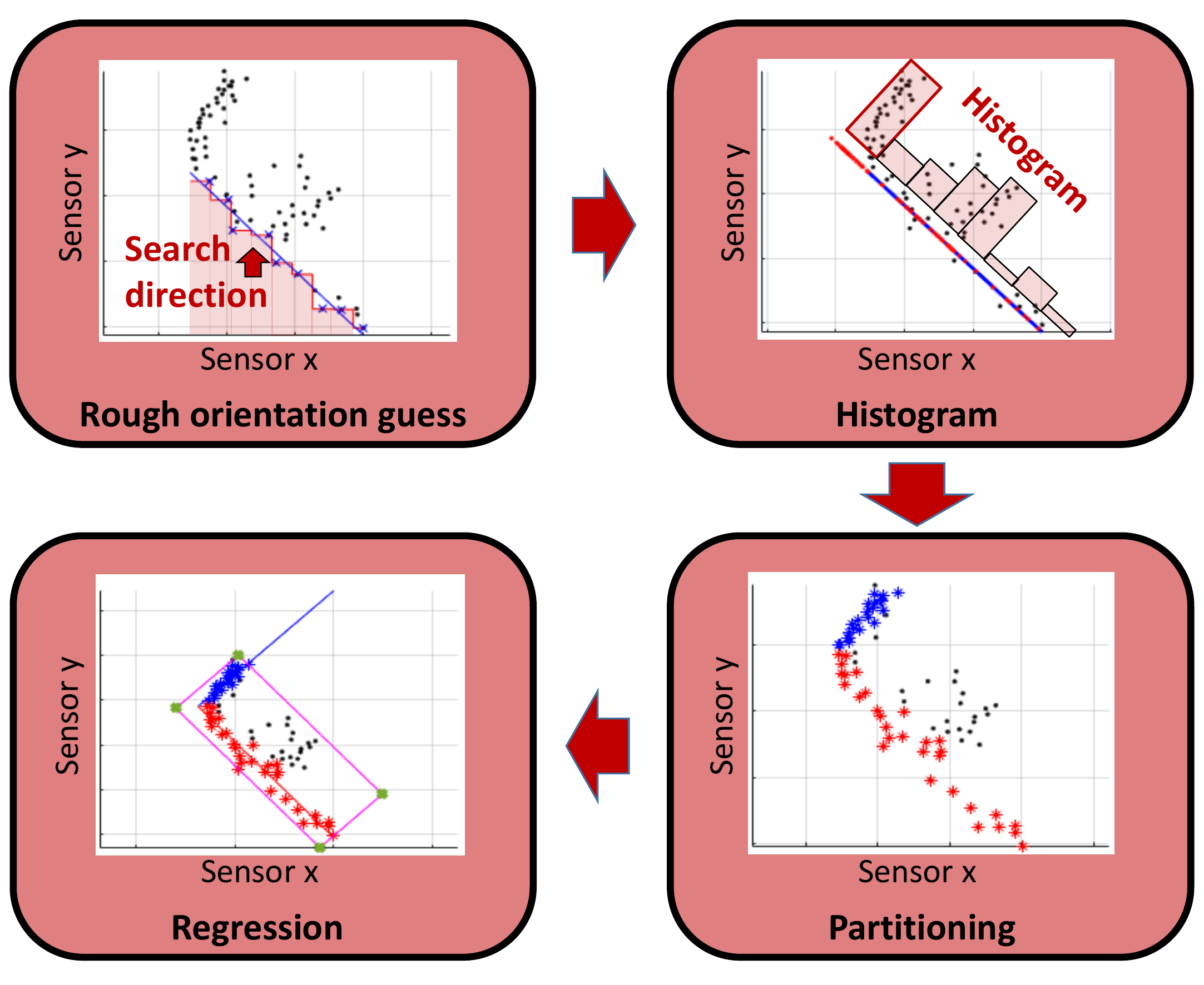}
	\caption{Based on a rough orientation estimation, a histogram is used to find the two partition of the L-shape. Finally a box is fitted by regression.}
	\label{fig:BoxFitting}
\end{figure}

Finally, the maximum bounding box along the estimated vehicle orientation is determined. This results in a box fit of the  CIV's shape. Fig. \ref{fig:BoxFitting} illustrates the algorithm steps of the box fitting.

\subsection{Model Matching}
In the next step, the information from the fitted bounding box in the sensor frame is matched with the cooperative information received from the CIV. Therefore, a sequence of measurements, where the CIV drives through the scene, using common reference points $\boldsymbol{p}$ of the CIV object, e.g. the box center, is evaluated.
In the first step, corresponding points in the sensor frame $\boldsymbol{p}_{j, \tilde{S}_i}$ and the world frame $\boldsymbol{p}_{j, w}$ are matched with respect to their time stamps.
As the CIV's measurements are not time synchronized with the measurements of the traffic monitoring system, the CIV's positions have to be interpolated over time.

Then, the yaw angle is determined, which aligns the transformed sensor frame $\tilde{\mathcal{F}}_i$ with the world frame $\mathcal{F}_w$. The vehicle position trace in the sensor and in the world frame are referred to the last point of the trace, respectively. Then, the mean difference angle $\beta_{\text{diff}}$ between the location vectors of corresponding positions is determined. As the influence of the position uncertainty to the angle estimation decreases with the length of the location vector, positions with a corresponding location vector length below a threshold $\theta_\beta$ are excluded from the difference angle estimation. This results in the rotation matrix
\begin{equation}
	\boldsymbol{R}_{\text{diff}} = \begin{bmatrix}
									\cos(\beta_{\text{diff}}) & -\sin(\beta_{\text{diff}}) \\
									\sin(\beta_{\text{diff}}) & \cos(\beta_{\text{diff}})\\
								\end{bmatrix}
\end{equation}
that aligns the yaw angle of the transformed sensor frame with the yaw angle of the world frame. Finally, the translation vector
\begin{equation}
	\boldsymbol{t}_{\tilde{S}_i}^{w} = \frac{1}{N_{\text{pos}}} \sum_{j=1}^{N_{\text{pos}}} \left( \boldsymbol{p}_{j, w} - \boldsymbol{R}_{\text{diff}} \; \boldsymbol{p}_{j, \tilde{S}_i} \right)
\end{equation}
is calculated. Thus,
\begin{equation}
	\small
	\mathcal{T}^{w}_{S_i} = \begin{bmatrix}
								\boldsymbol{R}_{\text{diff}} & \begin{matrix} 0 \\ 0 \end{matrix} & \boldsymbol{t}_{\tilde{S}_i}^{w} \\
								0 \; 0 & 1 & 0 \\
								0 \; 0 & 0 & 1
							\end{bmatrix} \; \cdot \; \mathcal{T}_{S_i}^{\tilde{S}_i}
\end{equation}
is the resulting overall transform.

\section{EVALUATION} \label{sec:Evaluation}
In this section, the experimental setup is described, the existence of systematic errors, e.g. a static offset in the CIV's position, is ruled out with a plausibility test, and the algorithm is evaluated with respect to the repetition error, i.e. the absolute value of the difference between the results of two independent calibration runs.

\subsection{Experimental Setup} \label{sec:Setup}
\begin{figure}[bt]
	\centering
	~\\[5pt]
	\includegraphics[width=0.25\textwidth]{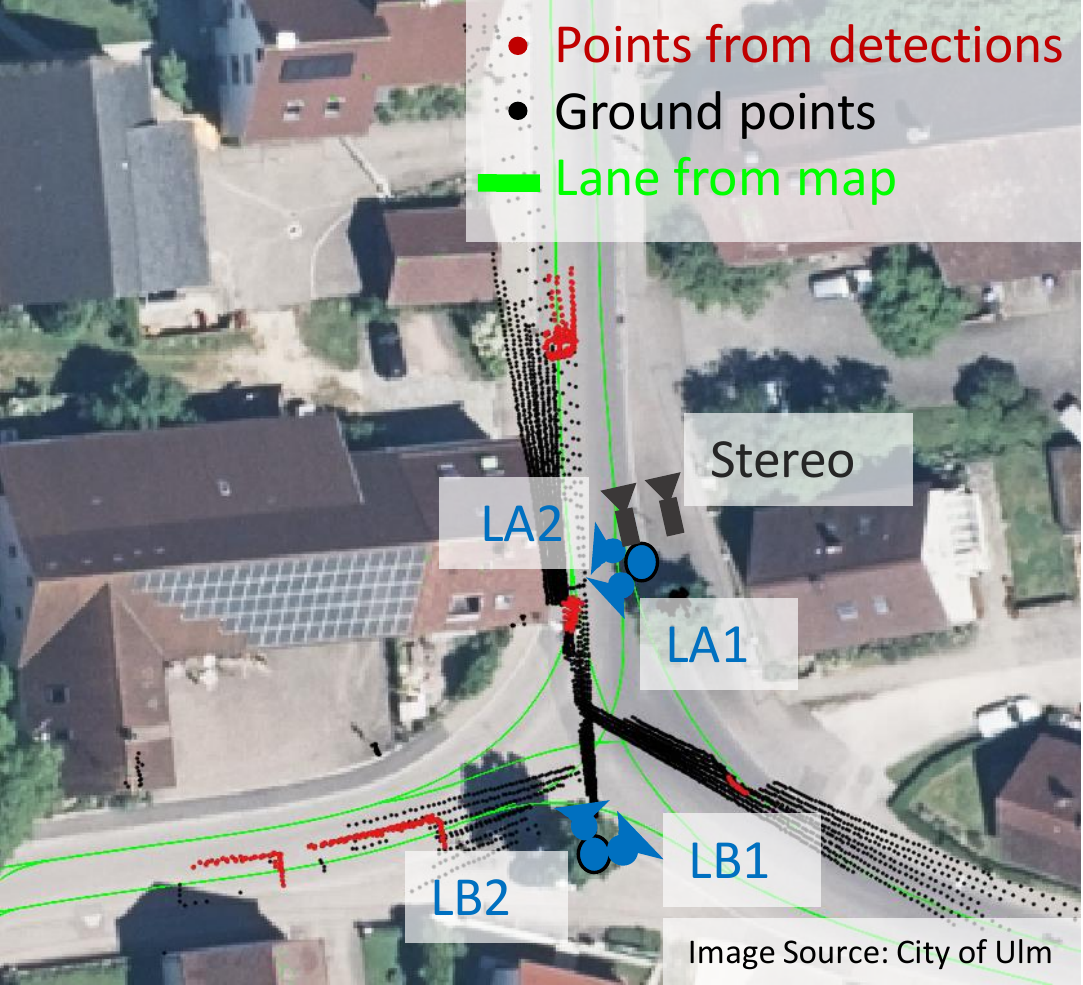}
	\caption{Set up of the four laserscanners LA1, LA2, LB1, LB2 as well as the stereo camera mapped into an aerial image. For better intuition, the calibration is already applied. Vehicles on the road are marked in red. Thus, a first plausibility check is possible.}
	\label{fig:LidarSetup}
\end{figure}

\begin{figure}[bt]
	\centering
	\includegraphics[width=0.25\textwidth]{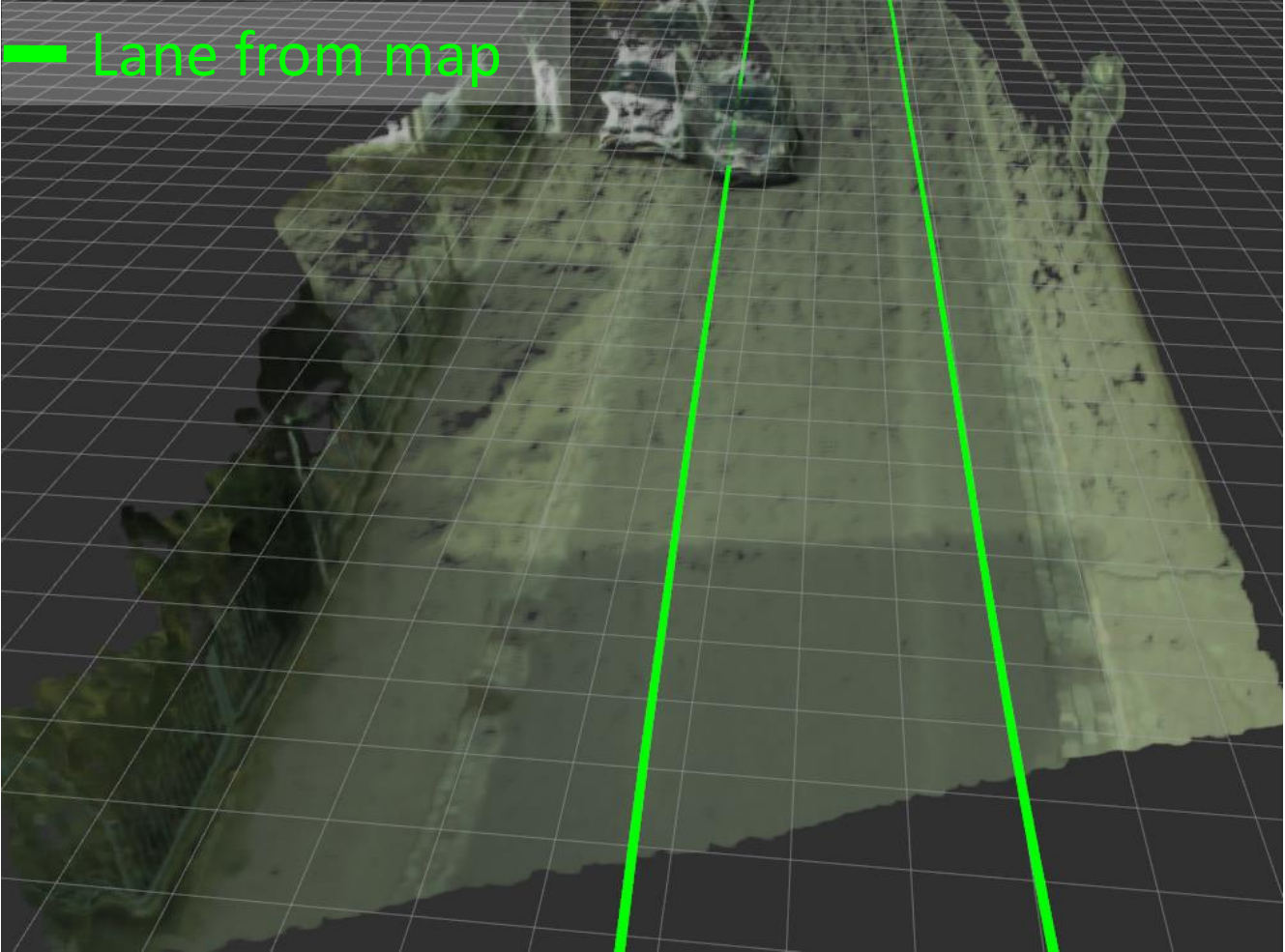}
	\caption{Visualization of the stero camera FOV. For better intuition, the calibration is already applied. Thus, a first plausibility check is possible.}
	\label{fig:StereoSetup}
\end{figure}

The algorithm has been applied to the calibration of a multi-sensor traffic monitoring system consisting of four SICK LDMRS 8 layer laserscaners and one stereo camera \cite{Buchholz2018}.
Figure \ref{fig:LidarSetup} shows ground points of the transformed laserscanner point cloud mapped into an aerial image. Furthermore, the position of the traffic monitoring system is sketched. It can be seen that except for the laserscanners LA2 and LB2, if at all, the sensors share a very small common FOV. Moreover, Fig. \ref{fig:LidarSetup} allows for a first plausibility check: as the ground points apparently lie on the street, the calculated calibration appears plausible.
To illustrate the FOV of the stereo vision system, the calibrated stereo image of the scene is depicted in Fig. \ref{fig:StereoSetup}. Again, the calibration appears plausible as the lanes from the map correspond well with the street in the stereo vision image.

 The laserscanners have a measurement uncertainty of $10 \, \text{cm}$ in range and an angular uncertainty of $0.5^\circ$. The depth resolution of the stereo camera system is depth dependent and thus can only be given approximately as $d \approx 60 \, \text{cm}$. The vehicle's position is determined using a high precision real time kinematic system, of which the measurement uncertainty is $2 \, \text{cm}$.

 \subsection{Exclusion of Systematic Errors}
\begin{figure}[bt]
 	\centering
 	\begin{tikzpicture}
 	\footnotesize
 	\input{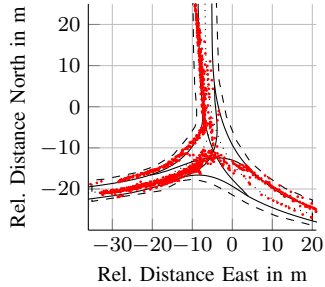}
 	\end{tikzpicture}	
 	\caption{Object position histogram over a $45 \,\text{min}$ recorded sequence. The reference lines are marked in solid black, while the street boarders are dashed black.}
 	\label{fig:SequenceHistogram}
 	\end{figure}

 To exclude the possibility of systematic errors that are unobservable within the repetition error used for further evaluation, the traffic is monitored for about $45 \, \text{min}$ resulting in approximately $12500$ object centers that are detected by an online object detection which was based on the calibration. These object centers are marked in the map of the monitored area. If there was a systematic error, the detected objects would appear at implausible positions. Fig. \ref{fig:SequenceHistogram} shows a map of the intersection as well as the recorded object centers and the reference line where vehicles are expected to appear. It can be seen that the detected objects match well with the reference line, thus it can be concluded that a remaining systematic error is improbable.

\subsection{Evaluation with respect to the Repetition Error}
The algorithm is tested on two sequences, each for all sensors, and then is evaluated with respect to reproduction error in the translation vector as well as in the Euler angles corresponding to the recovered rotation matrix. Table \ref{repetition_error} shows the repetition error in the Euler angles $|\Delta \psi_1|$, $|\Delta \psi_2|$, $|\Delta \psi_3|$ as well as in the translation $|\Delta x|$, $|\Delta y|$, $|\Delta z|$ for the multiple sensors.
\begin{table}[t]
	\footnotesize
	~\\[-27pt]
	\caption{Repetition error of calibration algorithm}
	\label{repetition_error}
	\begin{center}
		~\\[-19pt]
		\scriptsize
		\begin{tabular}{r|c|c|c|c|c|c}
			\hline
			Sensor & $|\Delta \psi_1|$ & $|\Delta \psi_2|$ & $|\Delta \psi_3|$ & $|\Delta x|$ & $|\Delta y|$ & $|\Delta z|$\\
			\hline
			LA1 & $0.06^{\circ}$  & $0.03^{\circ}$  & $0.07^{\circ}$ & 5.92 cm & 5.68 cm  & 0.04 cm\\
			\hline
			LA2 & $0.05^{\circ}$  & $0.01^{\circ}$  & $0.01^{\circ}$ & 1.75 cm & 8.47 cm  & 0.04 cm\\
			\hline
			LB1 & $0.13^{\circ}$  & $0.03^{\circ}$  & $0.003^{\circ}$ & 4.94 cm & 2.96 cm  & 0.7 cm\\
			\hline
			LB2 & $0.15^{\circ}$  & $0.01^{\circ}$  & $0.06^{\circ}$ & 12.9 cm & 7.42 cm  & 0.005 cm\\
			\hline
			Stereo & $0.05^{\circ}$  & $0.48^{\circ}$  & $0.18^{\circ}$ & 12.1 cm & 57.2 cm  & 1.41 cm\\			
			\hline
		\end{tabular}
	\end{center}
\end{table}
It has to be noted that a direct comparison to other calibration schemes \cite{Meissner2010}, \cite{Datondji2017}, \cite{Kuemmerle2018} proposed in literature so far proves difficult as the preconditions, the setup, and the evaluation method strongly differ through literature.
However, the maximum root mean squared error in translation $\Delta t_\text{max}$ and the maximum occurring angular error $\Delta \psi_\text{max}$ are common measures that could be calculated for all methods. Thus, these two measures are used for comparison.
This comparison is given in Table \ref{Comparison}.
\begin{table}[t] \label{Comparison}
	\footnotesize
	\caption{Comparison with literature results}
	\begin{center}
		~\\[-17pt]
		\scriptsize
		\begin{tabular}{r|c|c|c}
			\hline
			$ $ & Automation &$\Delta t_\text{max}$ & $\Delta \psi_\text{max}$ \\
			\hline
			Our approach (4 laserscanners) & full & $14.9 \, \text{cm}$ & $0,13^\circ$ \\
			\hline
			Datondji et al. \cite{Datondji2017} (1 fish-eye stereo) & semi & $15.3 \, \text{cm}$ & $2.9^\circ$ \\
			\hline
			Meissner et al. \cite{Meissner2010} & & & \\
			(special target, 4 laserscanners) & manual &  $2.7 \, \text{cm}$ & $0.3^\circ$ \\
			\hline	
			K{\"u}mmerle et al. \cite{Kuemmerle2018} (spherical target,$\,$ & & & \\
			 1 camera to 1 laserscanner) & manual & $0.5 \, \text{cm}$ & $0.07^\circ$ \\
			\hline	
		\end{tabular}
	\end{center}
\end{table}
The result of the semi-automatic method \cite{Datondji2017} is similar to ours, while the manual methods with special calibration target still outperform our calibration scheme in terms of accuracy while the necessary effort for these methods is significantly higher. As our results still are within the measurement uncertainty of our sensors, the calibration accuracy of our method is a good trade-off compared to the effortful manual methods.

\section{CONCLUSIONS} \label{sec:conclusion}

In this work, a low-effort self-calibration mechanism has been presented that uses the pose of a cooperative intelligent vehicle communicated by CAM messages to find the extrinsic calibration of a 3D imaging sensor. Thus, the sensors have been directly registered on a global frame and no overlapping FOV was needed. The algorithm has been evaluated for repetition accuracy for multiple laserscanners as well as for a stereo vision system showing sensor type independence as well as a repetition accuracy within the measurement uncertainty of the sensors.

In future work, the algorithm should be evaluated more extensively with an excessive series of repetition. Furthermore, the traffic monitoring system can be extended by further stereo vision sensors.

\addtolength{\textheight}{-12cm}   


\bibliographystyle{IEEEtran}
	{\footnotesize
		\bibliography{ITSC2019}}
\end{document}